\documentclass[apjl]{emulateapj}
\usepackage{graphicx}

\newcommand\kms{\ifmmode{\rm km\thinspace s^{-1}}\else km\thinspace s$^{-1}$\fi}
\newcommand\ms{\ifmmode{\rm m\thinspace s^{-1}}\else m\thinspace s$^{-1}$\fi}
\newcommand\msun{\ifmmode{M_{\odot}}\else $M_{\odot}$\fi}
\newcommand\rsun{\ifmmode{R_{\odot}}\else $R_{\odot}$\fi}
\newcommand\mjup{\ifmmode{M_{\rm Jup}}\else $M_{\rm Jup}$\fi}
\newcommand{\mysim}{\mathord{\sim}}
\newcommand{\reffigl}[1]{Figure~\ref{fig:#1}}
\newcommand{\refsecl}[1]{\mbox{Section \ref{sec:#1}}}

\newcommand{\reftabl}[1]{Table~\ref{tab:#1}}
\begin{document}

\shorttitle{T\lowercase{wo 'b's in the} B\lowercase{eehive}}
\shortauthors{Q\lowercase{uinn et al}.}

\title{T\lowercase{wo 'b's in the} B\lowercase{eehive}:
T\lowercase{he} D\lowercase{iscovery of the} F\lowercase{irst}
H\lowercase{ot} J\lowercase{upiters in an} O\lowercase{pen}
C\lowercase{luster}}

\author{
  Samuel N. Quinn\altaffilmark{1}, 
  Russel J. White\altaffilmark{1}, 
  David W. Latham\altaffilmark{2},
  Lars A. Buchhave\altaffilmark{3,4},
  Justin R. Cantrell\altaffilmark{1},
  Scott E. Dahm\altaffilmark{5},
  Gabor F\H{u}r\'esz\altaffilmark{2},
  Andrew H. Szentgyorgyi\altaffilmark{2},
  John C. Geary\altaffilmark{2},
  Guillermo Torres\altaffilmark{2},
  Allyson Bieryla\altaffilmark{2},
  Perry Berlind\altaffilmark{2},
  Michael C. Calkins\altaffilmark{2},
  Gilbert A. Esquerdo\altaffilmark{2},
  Robert P. Stefanik\altaffilmark{2}  
}

\altaffiltext{1}{Department of Physics \& Astronomy, Georgia State
  University, PO Box 4106, Atlanta, GA 30302}
\altaffiltext{2}{Harvard-Smithsonian Center for Astrophysics, 60
  Garden St, Cambridge, MA 02138}
\altaffiltext{3}{Niels Bohr Institute, University of Copenhagen,
  DK-2100 Copenhagen, Denmark}
\altaffiltext{4}{Centre for Star and Planet Formation, Natural History
  Museum of Denmark, University of Copenhagen, DK-1350 Copenhagen,
  Denmark}
\altaffiltext{5}{W.~M.~Keck Observatory, 65-1120 Mamalahoa Hwy,
  Kamuela, HI 96743}

\begin{abstract}

We report the discovery of two giant planets orbiting stars in
Praesepe (also known as the Beehive Cluster). These are the first
known hot Jupiters in an open cluster and the only planets known to
orbit Sun-like, main-sequence stars in a cluster. The planets are
detected from Doppler shifted radial velocities; line bisector spans
and activity indices show no correlation with orbital phase,
confirming the variations are caused by planetary companions. Pr0201b
orbits a $V=10.52$ late F dwarf with a period of $4.4264 \pm
0.0070$~days and has a minimum mass of $0.540 \pm 0.039~\mjup$, and
Pr0211b orbits a $V=12.06$ late G dwarf with a period of $2.1451 \pm
0.0012$~days and has a minimum mass of $1.844 \pm 0.064~\mjup$. The
detection of $2$ planets among $53$ single members surveyed
establishes a lower limit on the hot Jupiter frequency of
$3.8^{+5.0}_{-2.4}\%$ in this metal-rich open cluster. Given the
precisely known age of the cluster, this discovery also demonstrates
that, in at least $2$ cases, giant planet migration occurred within
$600$~Myr after formation. As we endeavor to learn more about the
frequency and formation history of planets, environments with
well-determined properties -- such as open clusters like Praesepe --
may provide essential clues to this end.

\end{abstract}


\section{Introduction}

Exoplanet studies over the last $15$ years have demonstrated that at
least $10\%$ of FGK stars harbor gas giant planets, with many of them
at surprisingly small separations, implying inward migration after
formation \citep{wright:2011}. Although the mechanism by which most
planets migrate is not understood, powerful constraints on proposed
theories of migration can be established by determining the orbital
properties of planets at young or adolescent ages ($<1$~Gyr). For
example, if migration occurs primarily due to interactions with a
circumstellar disk \citep[e.g.,][]{goldreich:1980,lin:1996}, the
migration must occur before the disk dissipates \citep[$\mysim
10$~Myr;][]{carpenter:2006}, and is predicted to circularize orbits.
Alternatively, if migration occurs primarily due to planet-planet
scattering \citep[e.g.,][]{adams:2003}, the process may take hundreds
of millions of years to occur and can produce highly eccentric orbits,
prior to any tidal circularization \citep[see review
by][]{lubow:2010}.

A direct way to find planets that can potentially be used to constrain
theories of migration is to search for them in young open
clusters. However, until now, only $2$ open cluster stars were known
to harbor planets -- $\epsilon$ Tau in the Hyades \citep{sato:2007}
and TYC 5409-2156-1 in NGC 2423 \citep{lovis:2007} -- both of which
are giant stars and thus, by necessity, have planets on wider orbits
than those occupied by hot Jupiters. In both cases the host stars are
of intermediate mass ($2.7$ and $2.4~\msun$), likely A or B type stars
when on the main sequence. The lack of detected planets orbiting FGK
main sequence stars (which are often referred to as Sun-like stars) in
open clusters has remained despite radial velocity (RV) suveys of $94$
dwarfs in the metal-rich Hyades \citep[][mean
${\rm[Fe/H]}=+0.13$]{paulson:2004} and $58$ dwarfs in M67
\citep{pasquini:2012}, as well as numerous transit searches in other
clusters
\citep[e.g.,][]{hartman:2009,pepper:2008,mochejska:2006}. While the
lack of detections may still be the result of small sample sizes
\citep[e.g.,][]{vansaders:2011}, millimeter-wave studies of disks
around stars in the Orion star forming region, which may evolve into
an open cluster, offer a plausible astrophysical
explanation. \citet{eisner:2008} find that most solar-type stars in
this region do not possess disks massive enough to form gas giant
planets. One may also speculate that for the few stars capable of
forming planets, the remaining disk masses may be insufficient to
permit inward migration \citep[see also][]{debes:2010}.

In an attempt to (1) more confidently determine whether planet
formation and/or migration is inhibited around stars within clusters,
and (2) potentially discover planets with known ages and measurable
orbital properties, we have carried out an RV survey of stars in the
Praesepe open cluster. Here we present the seminal result of that
survey - the discovery of the first $2$ hot Jupiters orbiting Sun-like
stars in a cluster.

\section{Sample Selection}
\label{sec:sample}

Stars were selected from the Praesepe open cluster because it is
relatively nearby ($170~{\rm pc}$), has $\mysim1000$ known members, a
well determined age \citep[600
Myr;][]{hambly:1995,kraus:2007,an:2007,gaspar:2009,delorme:2011}, and
significantly elevated metallicity (${\rm[Fe/H]} = +0.27 \pm 0.10$
dex, \citealt{pace:2008}; ${\rm[Fe/H]} = +0.11 \pm 0.03$,
\citealt{an:2007}). Its high metallicity is important because giant
planet frequency is strongly correlated with host star metallicity
\citep{santos:2004,fischer:2005,johnson:2010}; a metallicity as high
as $+0.27$ dex implies an increase in the giant planet frequency of a
factor of nearly $4$ relative to solar metallicity. If this
correlation applies to open cluster stars, as many as $1$ in $20$
Praesepe stars could harbor a hot Jupiter, and $1$ in $400$ could host
a transiting giant planet.

Cluster members were selected from the membership list assembled by
\citet{kraus:2007}, excluding stars with known spectroscopic or visual
companions \citep{mermilliod:2009,bouvier:2001,patience:2002}. To
ensure the velocity precision would be sufficient to detect substellar
companions, we limited our initial search to slowly rotating, bright
stars ($v \sin{i} < 12~\kms$; ${\rm V} < 12.3$). After applying these
cuts, our sample contained $65$ stars. Initial RV measurements
revealed $12$ stars to be obvious spectroscopic binaries ($\Delta{\rm
RV}>>1~\kms$) or non-members ($|{\rm RV} - {\rm RV_{cluster}}| >
5~\kms$), leaving $53$ viable targets for our search. These $53$ stars
are listed in \reftabl{stars}.

\begin{deluxetable}{lrrrrr}
  \tablewidth{0pc}
  \tablecaption{Target List and Observations Summary
    \label{tab:stars}
  }
  \tablehead{
    \colhead{Star} &
    \colhead{$\alpha$} &
    \colhead{$\delta$} &
    \colhead{V} &
    \colhead{N} &
    \colhead{$\sigma_{obs}$} \\
    \colhead{} &
    \colhead{(J2000)} &
    \colhead{(J2000)} &
    \colhead{(mag)} &
    \colhead{} &
    \colhead{(\ms)}
  }
  \startdata
  Pr0044 & $08:34:59.6$ & $+21:05:49.2$ & $11.06$ & $6$ & $42.0$ \\ 
  Pr0047 & $08:35:17.8$ & $+19:38:10.2$ & $12.24$ & $5$ & $13.4$ \\ 
  Pr0051 & $08:35:54.5$ & $+18:08:57.8$ & $10.88$ & $5$ & $18.3$ \\ 
  [-2.4ex]
  \enddata
  \tablecomments{
    \reftabl{stars} is presented in its entirety in the online
    journal. A portion is presented here for guidance regarding its
    form and content.
  }
\end{deluxetable}

\section{Observations}
\label{sec:obs}

We used the Tillinghast Reflector Echelle Spectrograph
\citep[TRES;][]{furesz:2008} mounted on the $1.5$-m Tillinghast
Reflector at the Fred L. Whipple Observatory on Mt. Hopkins, AZ to
obtain high resolution spectra of Praesepe stars, between UT
6-Jan-2012 and 16-Apr-2012. TRES is a temperature-controlled,
fiber-fed instrument with a resolving power of ${\rm R}\mysim44,000$
and a wavelength coverage of $\mysim 3850$-$9100$ \AA, spanning $51$
echelle orders.

We aimed to observe each target on two to three consecutive nights,
followed by another two to three consecutive nights $\mysim 1$~week
later. This strategy should be sensitive to most massive planets with
periods up to $10$~days. Though we were sometimes forced to deviate
from the planned observing cadence because of weather and instrument
availability, we were able to obtain $5$-$6$ spectra of each of our 53
targets. Exposure times ranged from $3$-$30$~minutes, yielding a
typical SNR per resolution element of $\mysim40$. We also obtained
nightly observations of the IAU RV standard star HD 65583, which is
$\mysim 14$~degrees from Praesepe. Precise wavelength calibration was
estaiblished by obtaining ThAr emission-line spectra before and after
each spectrum, through the same fiber as the science exposures.

\section{Analysis}
\label{sec:analysis}
\subsection{Spectroscopic Reduction and Cross Correlation}

Spectra were optimally extracted, rectified to intensity
vs. wavelength, and for each star the individual spectra were
cross-correlated, order by order, using the strongest exposure of that
star as a template \citep[for details, see][]{buchhave:2010}. We
typically used $\mysim 25$~orders, rejecting those plagued by telluric
absorption, fringing far to the red, and low SNR far to the blue. For
each epoch, the cross correlation functions (CCFs) from all orders
were added and fit with a Gaussian to determine the relative RV for
that epoch. Internal error estimates (which include, but may not be
limited to, photon noise) for each observation were calculated as
$\sigma_{int}={\rm RMS}(\vec{v})/\sqrt{N}$, where $\vec{v}$ is the RV
of each order, $N$ is the number of orders, and RMS denotes the
root-mean-squared velocity difference from the mean.

\begin{figure}[!htb]
\plotone{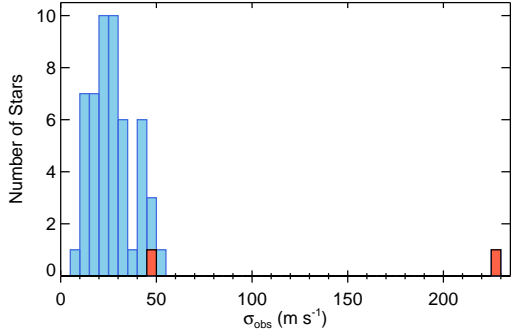}
\caption{ 
Observed velocity dispersions of the $53$ stars in our sample. The two
planet hosts are indicated by the solid red boxes. We note that while
Pr0201 resides in the upper tail of the distribution, other stars with
similar RMS values do not necessarily host planets; large internal
errors and larger than average jitter are two possible reasons for a
large RMS. For reference, a $1~\mjup$ planet in a $3$~day orbit around
a $1~\msun$ star has a full orbital amplitude of $281.7~\ms$.
\label{fig:rms}}
\end{figure}

To evaluate the significance of any potential velocity variation, we
compared the observed velocity dispersions ($\sigma_{obs}$),
illustrated in \reffigl{rms}, to the combined measurement
uncertainties, which we assumed stem from three sources: (1) internal
error, $\sigma_{int}$ (described above), (2) night-to-night
instrumental error, $\sigma_{\rm TRES}$, and (3) RV jitter induced by
stellar activity, $\sigma_*$.

Before assessing the instrumental error, we first used observations of
HD 65583 to correct for any systematic velocity shifts between runs
(such as a $25~\ms$ offset caused by a shutter upgrade in
mid-March). This was done by determining the median RV of HD 65583 for
each run, and adjusting all RVs from that run by the amount required
to make the median RV of HD 65583 constant over all runs. After
applying this correction, the RMS of the HD 65583 RVs was $10.8~\ms$,
with internal errors of only $6~\ms$. Since we expect negligibile
stellar jitter for the RV standard, we estimate the instrumental floor
error to be $\sigma_{\rm TRES}=\sqrt{10.8^2 - 6^2}~\ms = 9~\ms$.

In many cases the observed velocity dispersions are too large to be
explained by internal and instrumental errors alone, implying
substantial stellar jitter. We calculate the stellar jitter, $\sigma_*
= \sqrt{\sigma_{obs}^2 - \sigma_{int}^2 - \sigma_{\rm TRES}^2}$. The
mean stellar jitter is $13~\ms$, which is similar to that found by
\citeauthor{paulson:2004} ($16~\ms$) for coeval Hyades members.

Accounting for internal, instrumental, and stellar noise, we
constructed a $\chi^2$ fit of each star's RVs assuming a constant
velocity, and then calculated $P(\chi^2)$, the probability that the
observed $\chi^2$ value would arise from a star of constant RV. Pr0201
(BD+20 2184) and Pr0211 (2MASS J08421149+1916373) stood out, with
$P(\chi^2)<0.001$. We obtained additional spectra of these stars, and
in both cases a Lomb-Scargle periodogram revealed significant
periodicity. Their radial velocities are presented in \reftabl{rvs}.

\begin{deluxetable}{rrr|rrr}
  \tablewidth{0pc}
  \tablecaption{Relative Radial Velocities of Planet Hosts
    \label{tab:rvs}
  }
  \tablehead{
    \colhead{BJD} &
    \colhead{RV} &
    \colhead{$\sigma_{RV}$} &
    \colhead{BJD} &
    \colhead{RV} &
    \colhead{$\sigma_{RV}$} \\
    \colhead{($-2455900$)} &
    \multicolumn{2}{c}{(\ms)} &
    \colhead{($-2455900$)} &
    \multicolumn{2}{c}{(\ms)}
  }
  \startdata
  \multicolumn{6}{c}{Pr0201} \\
  \hline
  $ 32.841253$ & $-34.7$ & $18.6$ &   $ 94.824805$ & $ 19.0$ & $20.2$ \\
  $ 33.821938$ & $ 67.3$ & $23.9$ &   $ 96.733645$ & $139.4$ & $18.7$ \\
  $ 34.857994$ & $ 48.0$ & $21.8$ &   $ 97.744411$ & $-66.6$ & $17.7$ \\
  $ 39.945964$ & $-42.6$ & $20.2$ &   $ 98.649923$ & $-63.6$ & $29.6$ \\
  $ 41.020182$ & $-49.5$ & $27.0$ &   $121.701500$ & $ -7.3$ & $22.9$ \\
  $ 57.905296$ & $ -8.2$ & $21.4$ &   $122.646374$ & $ 81.9$ & $17.3$ \\
  $ 58.785164$ & $-29.3$ & $20.1$ &   $123.637192$ & $ 49.5$ & $16.1$ \\
  $ 59.818953$ & $  0.0$ & $12.0$ &   $124.646740$ & $ -6.3$ & $14.0$ \\
  $ 60.759119$ & $ 30.1$ & $17.0$ &   $125.693014$ & $ -4.1$ & $16.4$ \\
  $ 70.971372$ & $-18.7$ & $24.6$ &   $126.718364$ & $ 47.8$ & $30.4$ \\
  $ 71.796286$ & $-51.7$ & $42.5$ &   $128.683648$ & $  7.1$ & $18.3$ \\
  $ 72.970059$ & $ 13.5$ & $29.8$ &   $129.690867$ & $ 13.7$ & $18.4$ \\
  $ 81.771272$ & $  8.6$ & $23.5$ &   $130.724779$ & $ 89.9$ & $23.6$ \\
  $ 82.697156$ & $ 87.5$ & $15.9$ &   $131.705049$ & $ 85.2$ & $20.7$ \\
  $ 84.894042$ & $-90.5$ & $18.9$ &   $132.735355$ & $ 48.4$ & $24.8$ \\
  $ 86.805010$ & $ 27.0$ & $19.6$ &   $133.638578$ & $-18.2$ & $12.0$ \\
  $ 87.644768$ & $102.1$ & $19.6$ &      ~         &     ~   &    ~   \\
  \hline
  \multicolumn{6}{c}{Pr0211} \\
  \hline
  $ 86.772462$ & $ -75.7$ & $19.1$ &  $124.663763$ & $ 212.0$ & $21.4$ \\
  $ 87.720167$ & $ 413.8$ & $13.8$ &  $125.714331$ & $ 113.1$ & $16.9$ \\
  $ 88.721641$ & $-193.0$ & $21.3$ &  $126.737245$ & $ 275.7$ & $30.0$ \\
  $ 96.712491$ & $ 311.8$ & $24.9$ &  $128.698449$ & $ 428.7$ & $18.2$ \\
  $ 97.756832$ & $   0.0$ & $13.8$ &  $129.716113$ & $ -52.5$ & $21.0$ \\
  $ 98.662801$ & $ 405.8$ & $14.7$ &  $130.741942$ & $ 480.6$ & $26.5$ \\
  $121.718407$ & $ 346.4$ & $18.9$ &  $131.720488$ & $-156.9$ & $50.5$ \\
  $122.663113$ & $ 108.0$ & $20.3$ &  $132.715068$ & $ 504.4$ & $24.7$ \\
  $123.652317$ & $ 208.1$ & $21.8$ &  $133.654723$ & $ -98.1$ & $19.9$ \\ 
  \enddata
  \tablecomments{
    The errors listed here are internal error estimates, but in the
    orbital solutions we include an assumed stellar jitter of $13~\ms$
    and an instrumental floor error of $9~\ms$, added in quadrature
    with the internal errors.
  }
\end{deluxetable}

\subsection{Orbital Solutions}
\label{sec:orbits}

We used a Markov Chain Monte Carlo (MCMC) analysis to fit Keplerian
orbits to the radial velocity data of Pr0201 and Pr0211, fitting for
orbital period $P$, time of conjunction $T_{\rm c}$, the radial
velocity semi-amplitude $K$, the center-of-mass velocity $\gamma$, and
the orthogonal quantities $\sqrt{e}\cos{\omega}$ and
$\sqrt{e}\sin{\omega}$, where $e$ is eccentricity and $\omega$ is the
argument of periastron. We calculated errors from the extent of the
central $68.3\%$ interval of the MCMC posterior distributions.

The full orbital solutions give eccentricities of $e =
0.156^{+0.041}_{-0.112}$ for Pr0201 and $e = 0.046^{+0.021}_{-0.024}$
for Pr0211. However, it can take many precise observations to
accurately measure small, non-zero eccentricities
\citep[e.g.,][]{zakamska:2011}, and both are consistent with $e=0$ to
within $2$-$\sigma$, so we advise caution to not over-interpret these
results; for short period planets such as these, we expect that in the
absence of additional bodies, tidal forces should have already
circularized the orbits \citep[e.g.,][]{adams:2006}. We also note that
the other orbital parameters are changed by less than $1$-$\sigma$
when fixing $e=0$, so in the absence of additional data, the
assumption of circularized orbits is acceptable. We report the
solutions with $e=0$ in \reftabl{props} and plot the best fit circular
orbits in \reffigl{orbits}.

\subsection{Line Bisectors and Stellar Activity Indices}

\begin{figure*}[!htb]
\plottwo{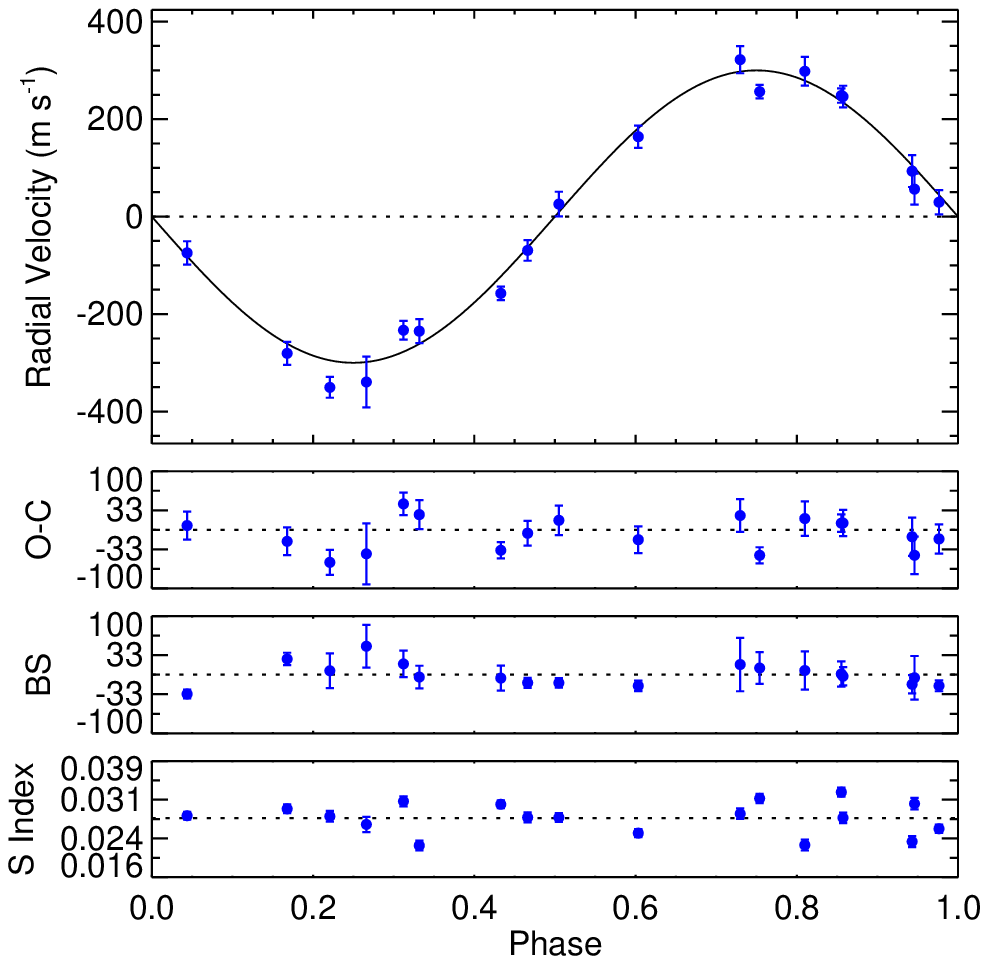}{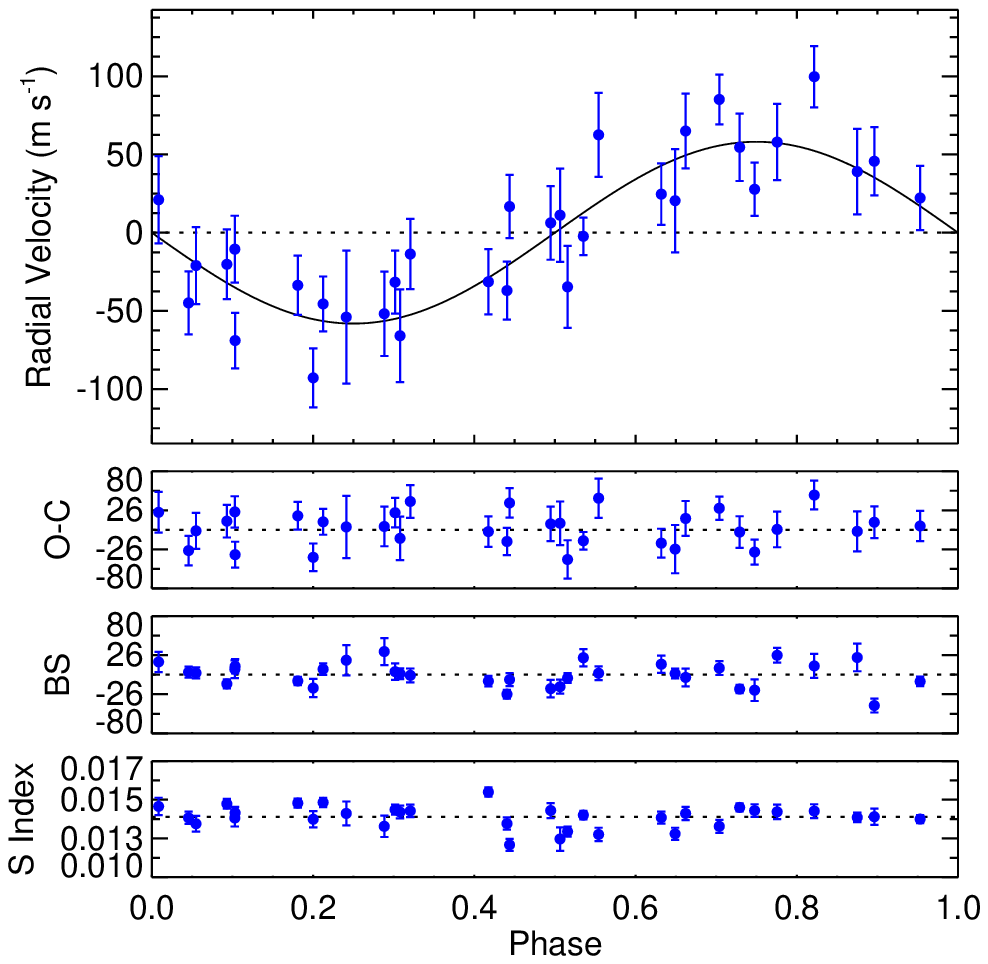}
\caption{ 
Orbital solutions for Pr0211 (left) and Pr0201 (right). The panels,
from top to bottom, show the relative RVs, best-fit residuals,
bisector span variations, and relative S index values. RV error bars
represent the internal errors, and do not include astrophysical
jitter, although $13~\ms$ jitter was assumed in the orbital fit. The
solid curve shows the best-fit orbital solution. The orbital
parameters are listed in \reftabl{props}.
\label{fig:orbits}}
\end{figure*}

If the observed velocity variations were caused by a background blend
\citep{mandushev:2005} or star spots \citep{queloz:2001}, we would
expect the shape of a star's line bisector to vary in phase with the
radial velocities. A standard prescription for characterizing the
shape of a line bisector is to measure the difference in relative
velocity of the top and bottom of a line bisector; this difference is
referred to as a line bisector span \citep[see,
e.g.,][]{torres:2005}. To test against background blends or star
spots, we computed the line bisector spans for all observations of
Pr0201 and Pr0211. As illustrated in \reffigl{orbits}, the bisector
span variations are small ($\sigma_{BS} < 20~\ms$) and are not
correlated with the observed RV variations. As an additional check
against activity induced RV variations, for each spectrum of Pr0201
and Pr0211 we also compute the S index -- an indicator of
chromospheric activity in the CaII H\&K lines. We follow the procedure
of \citet{vaughan:1978}, but we note that our S indices are not
calibrated to their scale; these are relative measurements. As shown
in \reffigl{orbits}, there is no correlation with orbital phase. These
line bisector and S index comparisons strongly support the conclusion
that the observed RV variations are caused by planetary companions.

\subsection{Stellar and Planetary Properties}
\label{sec:props}

We used the spectroscopic classification technique Stellar Parameter
Classification \citep[SPC;][]{buchhave:2012} to determine effective
temperature $T_{\rm eff}$, surface gravity $\log{g}$, projected
rotational velocity $v\sin{i}$, and metallicity [m/H] for each of our
target stars. In essence, SPC cross correlates an observed spectrum
against a grid of synthetic spectra, and uses the correlation peak
heights to fit a $3$-dimensional surface in order to find the best
combination of atmospheric parameters ($v\sin{i}$ is fit iteratively
since it is only weakly correlated to changes in the other
parameters). We used the CfA library of synthetic spectra, which are
based on Kurucz model atmospheres \citep{kurucz:1992} calculated by
John Laird for a linelist compiled by Jon Morse. Like other
spectroscopic classification techniques, SPC can be limited by
degeneracy between parameters, notably $T_{\rm eff}$, $\log{g}$, and
[m/H], but in this case we can enforce the known cluster metallicity
to partially break that degeneracy. Though we leave detailed
description of the ensemble sample for a subsequent paper, from an
analysis of the $53$ stars in our sample, we calculated a cluster
metallicity of ${\rm [m/H]}=+0.187 \pm 0.038$. This value is
consistent with previous estimates \citep[e.g., $+0.27 \pm 0.10$,
][]{pace:2008}. 

Regarding the planet hosts Pr0201 and Pr0211 in particular, we note
that our derived temperatures of $6174$ K and $5326$ K are in
agreement with published spectral types \citep[F7.5 and
G9.3;][]{kraus:2007}, and that their individual SPC-derived
metallicities ($+0.18 \pm 0.08$ and $+0.19 \pm 0.08$) are consistent
with the median of the cluster. Detailed heavy-element abundance
analyses of Pr0201 are reported in \citet{pace:2008} and
\citet{maiorca:2011}.

We used the stellar parameters from SPC and the known age of Praesepe
in conjunction with the Yonsei-Yale stellar models \citep{yi:2001} to
extract the stellar masses and radii. We found that the $\log{g}$
values indicated by the isochrone fits were slightly more than
$1$-$\sigma$ lower than the SPC values, but it is possible that the
formal errors for SPC are too small and/or that the stellar models are
inaccurate for these somewhat young and active stars. We iterated the
SPC analysis, this time fixing both the cluster metallicity and the
$\log{g}$ from the isochrone fit. This resulted in a slightly lower
$T_{\rm eff}$, and the subsqequent isochrone fit was consistent with
all of the SPC parameters. Using an age of $578$~Myr
\citep{delorme:2011}, we adopted stellar masses and radii from the
isochrone fits ($M_* = 1.234 \pm 0.034~\msun$, $R_* = 1.167 \pm
0.121~\rsun$ for Pr0201; $M_* = 0.952 \pm 0.040~\msun$, $R_* = 0.868
\pm 0.078~\rsun$ for Pr0211), but caution that the formal errors on
stellar and planetary masses and radii do not encompass any potential
systematics. The estimates of the stellar masses provide lower limits
on the masses of the planets of $0.540 \pm 0.039~\mjup$ for Pr0201b
and $1.844 \pm 0.064~\mjup$ for Pr0211b. \reftabl{props} lists all of
the stellar and planetary properties.

\begin{deluxetable}{lcc}
  \tablewidth{0pc} \tablecaption{Stellar and Planetary Properties
    \label{tab:props}
  }
  \tablehead{
    \colhead{} &
    \colhead{Pr0201} &
    \colhead{Pr0211}
  }
  \startdata
  \multicolumn{3}{l}{Orbital Parameters}                                      \\
  \hline
  $P$ [days]            & $4.4264 \pm 0.0070$     & $2.1451 \pm 0.0012$       \\
  $T_{\rm c}$ [BJD]     & $2455992.861 \pm 0.053$ & $2456013.9889 \pm 0.0072$ \\
  $K$ [\ms]             & $58.1 \pm 4.1$          & $299.9 \pm 6.1$           \\
  $e$                   & $0$                     & $0$                       \\
  $\gamma$ [\kms]       & $34.035 \pm 0.101$      & $35.184 \pm 0.198$        \\
  \hline
  \multicolumn{3}{l}{Stellar and Planetary Properties}                        \\
  \hline
  $M_*$ [\msun]         & $1.234 \pm 0.034$       & $0.952 \pm 0.040$       \\
  $R_*$ [\rsun]         & $1.167 \pm 0.121$       & $0.868 \pm 0.078$       \\
  $T_{\rm eff,*}$ [K]\tablenotemark{a}
                        & $6174 \pm 50$           & $5326 \pm 50$           \\
  $\log{g}_*$ [dex]\tablenotemark{a}
                        & $4.41 \pm 0.10$         & $4.55 \pm 0.10$         \\
  $v\sin{i}$ [\kms]     & $9.6 \pm 0.5$           & $4.8 \pm 0.5$           \\
  $[m/H]$ [dex]\tablenotemark{a}
                        & $0.187 \pm 0.038$       & $0.187 \pm 0.038$       \\
  $Age$ [Myr]\tablenotemark{b}
                        & $578 \pm 49$            & $578 \pm 49$            \\
  $M_p \sin{i}$ [\mjup] & $0.540 \pm 0.039$       & $1.844 \pm 0.064$       \\
  [-2.4ex]
  \enddata
  \tablenotetext{a}{
    From the final SPC iteration. [m/H] was fixed to the mean cluster
    metallicity calculated from an SPC analysis of our $53$ stars. See
    \refsecl{props}.
  }
  \tablenotetext{b}{
    From \citet{delorme:2011}.
  }
  \tablecomments{
    The orbital parameters correspond to the best fit circular orbit
    (see \refsecl{orbits}).
  }
\end{deluxetable}

\section{Discussion}
\label{sec:disc}

Our discovery of two hot Jupiters in Praesepe confirms that
short-period planets do exist in open clusters. Moreover, assuming
these gas giants formed beyond the snow-line, the planets have
migrated to nearly circular short period orbits in $600$~Myr. Although
a more complete analysis that takes into account the detection limits
of our entire sample is called for, we can already place some
constraints on the hot Jupiter frequency in Praesepe. If we make the
assumption that the observations of the other $51$ stars in our sample
can completely rule out the presence of short-period, massive planets,
then we obtain a lower limit on the hot Jupiter frequency in Praesepe:
$(2^{+2.6}_{-1.3})/53$; at least $3.8^{+5.0}_{-2.4}\%$ of all single
FGK cluster members host a hot Jupiter \citep[Poisson errors were
calculated following the prescription in][]{gehrels:1986}. While this
number is slightly higher than the frequency for field stars
\citep[$1.20 \pm 0.38 \%$;][]{wright:2012}, it is consistent with that
expected from the enriched metallicity environment of Praesepe.

Uncertainties in planetary properties are most often limited by
determination of properties of their host stars, but planets in
clusters -- particularly those that transit their host stars -- can
yield greatly reduced observational uncertainties. The observable
transit parameter $a/R_*$ constrains the stellar $\log{g}$
\citep{sozzetti:2007}, and the cluster's mean metallicity can be
determined more precisely than that of any one star. Accurate
$\log{g}$ and [m/H] values will also improve the spectroscopic $T_{\rm
eff}$ estimates by breaking the degeneracy between the three
parameters. When combined with the cluster age and distance, the
resulting range of allowed masses and radii from stellar models would
be greatly reduced. The precision in the stellar properties would
propagate to an extremely precise planetary mass, radius, and age,
providing a better test for models of planetary structure and
evolution. Just as they have played an important role in the
calibration of stellar properties, open clusters hold great promise as
laboratories to explore properties of exoplanets at various ages and
with great precision.

\acknowledgements 
This material is based upon work supported by the National Aeronautics
and Space Administration (NASA) under Grant No. NNX11AC32G issued
through the Origins of Solar Systems program. D.~W.~L. acknowledges
partial support from NASA's Kepler mission under Cooperative Agreement
NNX11AB99A with the Smithsonian Astrophysical Observatory.

{\it Facility:} \facility{FLWO:1.5m (TRES)}


\begin{thebibliography}{}

\bibitem[Adams \& Laughlin(2003)]{adams:2003} Adams, F.~C. \&
  Laughlin, G.~2003, Icarus, 163, 290

\bibitem[Adams \& Laughlin(2006)]{adams:2006} Adams, F.~C. \&
  Laughlin, G.~2006, \apj, 649, 1004

\bibitem[An et al.(2007)]{an:2007} An D., Terndrup, D.~M., \&
  Pinsonneault, M.~H.~2007, \apj, 671, 1640

\bibitem[Bouvier et al.(2001)]{bouvier:2001} Bouvier, J., Duchene, G.,
  Mermilliod, J.~C., \& Simon, T.~2001, \aap, 375, 989

\bibitem[Bressert et al.(2011)]{bressert:2011} Bressert E., Bastian,
  N., \& Gutermuth, R.~2011, JENAM conference, arXiv:1102.0565

\bibitem[Buchhave et al.(2010)]{buchhave:2010} Buchhave, L.~A., et
  al.~2010, \apj, 720, 1118

\bibitem[Buchhave et al.(2012)]{buchhave:2012} Buchhave, L.~A., et
  al.~2012, Nature, 486, 375

\bibitem[Carpenter et al.(2006)]{carpenter:2006} Carpenter, J.~M.,
  Mamajek, E.~E., Hillenbrand, L.~A., \& Meyer, M.~R.~2006, \apj, 651,
  49

\bibitem[Debes \& Jackson(2010)]{debes:2010} Debes, J.~H. \& Jackson,
  B.~2010, \apj, 723, 1703

\bibitem[Delorme et al.(2011)]{delorme:2011} Delorme, P., et al.~2011,
  \mnras, 413, 2218

\bibitem[Eisner et al.(2008)]{eisner:2008} Eisner, J.~A., Plambeck,
  R.~L., Carpenter, J.~M., Corder, S.~A., Qi, C., \& Wilner, D.~2008,
  \apj, 683, 304

\bibitem[Fischer \& Valenti(2005)]{fischer:2005} Fischer, D.~A. \&
  Valenti, J.~2005, \apj, 622, 1102

\bibitem[F\H{u}r\'esz(2008)]{furesz:2008} F\H{u}r\'esz, G.~2008,
  Ph.D.~thesis, University of Szeged, Hungary

\bibitem[Gehrels(1986)]{gehrels:1986} Gehrels, N.~1986, \apj, 303, 336

\bibitem[G\'asp\'ar et al.(2009)]{gaspar:2009} G\'asp\'ar, A., et
  al.~2009, \apj, 697, 1578

\bibitem[Goldreich \& Tremaine(1980)]{goldreich:1980} Goldreich, P. \&
  Tremaine, S.~1980, \apj, 241, 425

\bibitem[Hambly et al.(1995)]{hambly:1995} Hambly, N.~C., Steele,
  I.~A., Hawkins, M.~R.~S., \& Jameson, R.~F.~1995, \mnras, 273, 505

\bibitem[Hartman et al.(2009)]{hartman:2009} Hartman, J.~D., et
  al.~2009, \apj, 695, 336

\bibitem[Johnson et al.(2010)]{johnson:2010} Johnson, J.~A., Aller,
  K.~M., Howard, A.~W., \& Crepp, J.~R.~2010, \pasp, 122, 905

\bibitem[Kraus \& Hillenbrand(2007)]{kraus:2007} Kraus, A.~L. \&
  Hillenbrand, L.~A.~2007, \aj, 134, 2340

\bibitem[Kurucz(1992)]{kurucz:1992} Kurucz, R.~L.~1992, IAUS, 149, 225

\bibitem[Lin et al.(1996)]{lin:1996} Lin, D.~N.~C., Bodenheimer, P.,
  \& Richardson, D.~C.~1996, Nature, 380, 606

\bibitem[Lovis \& Mayor(2007)]{lovis:2007} Lovis, C. \& Mayor,
  M.~2007, \aap, 472, 657

\bibitem[Lubow \& Ida(2010)]{lubow:2010} Lubow, S.~H., \& Ida,
  S.~2010, in Exoplanets, ed.~S.~Seager (Tucson, AZ: Univ.~Arizona
  Press), 347

\bibitem[Maiorca et al.(2011)]{maiorca:2011} Maiorca, E., et al.~2011,
  \apj, 736, 120

\bibitem[Mandushev et al.(2005)]{mandushev:2005} Mandushev, G., et
  al.~2005, \apj, 621, 1061

\bibitem[Mermilliod et al.(2009)]{mermilliod:2009} Mermilliod, J.-C.,
  Mayor, M., \& Udry, S.~2009, \aap, 498, 949

\bibitem[Mochejska et al.(2006)]{mochejska:2006} Mochejska, B.~J., et
  al.~2006, \aj, 131, 1090

\bibitem[Pace et al.(2008)]{pace:2008} Pace, G., Pasquini, L., \&
  Francois, P.~2008, \aap, 489, 403

\bibitem[Pasquini et al.(2012)]{pasquini:2012} Pasquini, L., et
  al.~2012, \aap, in press (arXiv:1206.5820).

\bibitem[Patience et al.(2002)]{patience:2002} Patience, J., Ghez,
  A.~M., Reid, I.~N., \& Matthews, K.~2002, \aj, 123, 1570

\bibitem[Paulson et al.(2004)]{paulson:2004} Paulson, D.~B., Cochran,
  W.~D., \& Hatzes, A.~P.~2004, \aj, 127, 3579

\bibitem[Pepper et al.(2008)]{pepper:2008} Pepper, J., et al.~2008,
  \aj, 135, 907

\bibitem[Queloz et al.(2001)]{queloz:2001} Queloz, D., et al.~2001,
  \aap, 379, 279

\bibitem[Santos et al.(2004)]{santos:2004} Santos, N.~C., Israelian,
  G., \& Mayor, M.~2004, Planetary Systems in the Universe, 202, 118

\bibitem[Sato et al.(2007)]{sato:2007} Sato, B., et al.~2007, \apj,
  661, 527

\bibitem[Sozzetti et al.(2007)]{sozzetti:2007} Sozzetti, A., Torres,
  G., Charbonneau, D., Latham, D.~W., et al.~2007, \apj, 664, 1190

\bibitem[Torres et al.(2005)]{torres:2005} Torres, G., Konacki, M.,
  Sasselov, D.~D., \& Jha, S.~2005, \apj, 619, 558

\bibitem[van Saders \& Gaudi(2011)]{vansaders:2011} van Saders,
  J.~L. \& Gaudi, B.~S.~2011, \apj, 729, 63

\bibitem[Vaughan et al.(1978)]{vaughan:1978} Vaughan, A.~H., Preston,
  G.~W., \& Wilson, O.~C.~1978, \pasp, 90, 267

\bibitem[Wright et al.(2011)]{wright:2011} Wright, J.~T., et al.~2011,
  \pasp, 123, 412

\bibitem[Wright et al.(2012)]{wright:2012} Wright, J.~T., et al.~2012,
  \apj, in press (arXiv:1205.2273).

\bibitem[Yi et al.(2001)]{yi:2001} Yi, S., et al.~2001, ApJS, 136,
  417

\bibitem[Zakamska et al.(2011)]{zakamska:2011} Zakamska, N.~L., Pan,
  M., \& Ford, E.~B.~2011, \mnras, 410, 1895

\end{thebibliography}
\end{document}